**Title:**

The availability of research data declines rapidly with article age.

**Authors:** Timothy H. Vines [1,2,*], Arianne Y.K. Albert [3], Rose L. Andrew [1], Florence Débarre [1,4], Dan G. Bock [1], Michelle T. Franklin [1,5], Kimberly J. Gilbert [1], Jean-Sébastien Moore [1,6], Sébastien Renaut [1], Diana J. Rennison [1]

**Affiliations:** [1]Biodiversity Research Centre, University of British Columbia, 6270 University Blvd Vancouver BC, Canada, V6T 1Z4. [2]Molecular Ecology Editorial Office, 6270 University Blvd Vancouver BC, Canada, V6T 1Z4. [3]Women's Health Research Institute, 4500 Oak Street, Vancouver, BC, Canada V6H 3N1. [4]Centre for Ecology & Conservation Biosciences, University of Exeter, Cornwall Campus, Tremough, Penryn, TR10 9EZ, UK. [5]Institute for Sustainable Horticulture, Kwantlen Polytechnic University, 12666 72nd Avenue, Surrey, British Columbia, Canada, V3W 2M8. [6]Department of Biology, Université Laval, 1030 Avenue de la Médecine, Laval, QC G1V 0A6

[*]Author for correspondence: vines@zoology.ubc.ca. Tel: 1 778 989 8755 Fax: 1 604 822 8982

**Running Head**: Data availability declines with article age

**Highlights:**

- We examined the availability of data from 516 studies between 2 and 22 years old.
- The odds of a dataset being reported as extant fell by 17% per year.
- Broken emails and obsolete storage devices were the main obstacles to data sharing.
- Policies mandating data archiving at publication are clearly needed.

**Summary**

Policies ensuring that research data are available on public archives are increasingly being implemented at the government [1], funding agency [2-4], and journal [5,6] level. These policies are predicated on the idea that authors are poor stewards of their data, particularly over the long term [7], and indeed many studies have found that authors are often unable or unwilling to share their data [8-11]. However, there are no systematic estimates of how the availability of research data changes with time since publication. We therefore requested datasets from a relatively homogenous set of 516 articles published between 2 and 22 years ago, and found that availability of the data was strongly affected by article age. For papers where the authors gave the status of their data, the odds of a dataset being extant fell by 17% per year. In addition, the odds that we could find a working email address for the first, last or corresponding author fell by 7% per year. Our results reinforce the notion that, in the long term, research data cannot be reliably preserved by individual researchers, and further demonstrate the urgent need for policies mandating data sharing via public archives.

## Results

We investigated how research data availability changes with article age. To avoid potential confounding effects of data type and different research community practices, we focused on recovering data from articles containing morphological data from plants or animals that made use of a Discriminant Function Analysis (DFA). Our final dataset consisted of 516 articles published between 1991 and 2011. We found at least one apparently working email for 385 papers (74%), either in the article itself or by searching online. We received 101 datasets (19%), and were told that another 20 (4%) were still in use and could not be shared, such that a total of 121 datasets (23%) were confirmed as extant. Table 1 provides a breakdown of the data by year.

We used logistic regression to formally investigate the relationships between the age of the paper and 1) the probability that at least one email appeared to work (i.e. did not generate an error message); 2) the conditional probability of a response given that at least one email appeared to work; 3) the conditional probability of getting a response that indicated the status of the data (data lost, exist but unwilling to share, or data shared) given that a response was received; and finally 4) the conditional probability the data was extant (either 'shared' or 'exists but unwilling to share') given that an informative response was received.

There was a negative relationship between the age of the paper and the probability of finding at least one apparently working email either in the paper or by searching online (OR = 0.93 [0.90 – 0.96, 95% CI], $p$-value < 0.00001). The odds ratio suggests that for every year since publication, the odds of finding at least one apparently working email decreased by 7% (Figure 1A). Since we searched for emails in both the paper and online, four factors contribute to the probability of finding a working email: i) the number of emails in the paper and ii) the chance that any of those worked, iii) the number of emails we could find by searching online and iv) the chance that any of those worked. The total number of email addresses we found in the paper did decrease with age (Poisson regression coefficient = -0.07, SE = 0.01, $p$-value < 0.0001) from an average of 1.17 in 2011 to 0.42 in 1991 (Figure 2A), and there was a slight positive effect of article age on the number of emails we found online (Poisson regression coefficient = 0.015, SE = 0.007, $p$-value <0.05, Figure 2C). Moreover, the chance that an email found in the paper or online appeared to work also showed a relationship with article age (OR = 0.96 [0.926 – 0.998, 95% CI], $p$-value < 0.05, and OR = 0.97 [0.936 – 0.997, 95% CI], $p$-value < 0.05, respectively), such that the odds that an email appeared to work declined by 4% and 3% per year since publication, respectively (Figure 2B and 2D).

We note that eight email addresses generated an error message but did lead to a response from the authors. It also seems likely that some addresses failed but did not generate an error message, leading us to record a 'no response' rather than 'email not working', although unfortunately the frequency of these cannot be estimated from our data.

There was no relationship between age of the paper and the probability of a response given that there was an apparently working email (50% response rate, OR = 1.00 [0.97 – 1.04, 95% CI], Figure 1B). There was also no relationship between article age and the probability that the response indicated the status of the data, given a response was received (83% useful responses, OR = 1.00 [0.95 – 1.07, 95% CI], Figure 1C).

Finally, there was a strong negative relationship between the age of the paper and the probability that the dataset was still extant (either 'shared' or 'exists but unwilling to share'), given that a response indicating the status of the data was received (OR = 0.83 [0.79 – 0.90, 95% CI], $p$-value < 0.0001, Figure 1D). The odds ratio suggests that for every yearly increase in article age, the odds of the dataset being extant decreased by 17%.

## Discussion

We found a strong effect of article age on the availability of data from these 516 studies. The decline in data availability could arise because the authors of older papers were less likely to respond, but this was not supported by the data. Instead, researchers were equally likely to respond (Figure 1B) and to indicate the status of their data (Figure 1C) across the entire range of article ages.

The major cause of the reduced data availability for older papers was the rapid increase in the proportion of datasets reported as either lost or on inaccessible storage media. For papers where authors reported the status of their data, the odds of the data being extant decreased by 17% per year (Figure 1D). There was a continuum of author responses between the data being reported lost and being stored on inaccessible media, and seemed to vary with the amount of time and effort involved in retrieving the data. Responses ranged between authors being sure that the data were lost (e.g. on a stolen computer), thinking they might be stored in some distant location (e.g. their parent's attic), or having some degree of certainty that the data are on a Zip or floppy disk in their possession but they no longer have the appropriate hardware to access it. In the latter two cases, the authors would have to devote hours or days to retrieving the data. Our reason for needing the data (a reproducibility study) was not especially compelling for authors, and we may have received more of these inaccessible datasets if we had offered authorship on the subsequent paper, or said that the data were needed for an important medical or conservation project.

The odds that we were able to find an apparently working email address (either in the paper or by searching online) for any of the contacted authors did decrease by about 7% per year. This decrease was partly driven by a dearth of email addresses in articles published before 2000 (0.38 per paper on average for 1991-1999) compared with those published after 2001 (1.08 per paper on average, Figure 2A). Wren et al. [12] found a similar increase in the number of emails in articles published after 2000. The larger number of emails in recent papers may mean that the issue of missing author emails is restricted to articles from before 2000: researchers in e.g. 2031 will be able to try a wider range of addresses in their attempts to contact authors of articles published in 2011.

The proportion of emails from the paper that appeared to work declined with article age between 2 and 14 years of age, and then rose to around 80% for articles from 1991, 1993 and 1995 (Figure 2B). These latter three proportions are only based on a total of 13 email addresses. Wren et al. [12] reported a steep decline with age in the proportion of functioning emails from papers published between 1995 and 2004, such that 84% of their ten-year-old emails returned an error message. Our proportions for ten-year-old emails are lower, with only 51% of emails from 2003 returning an error. It may be that email addresses are becoming more stable through time, although this clearly requires additional study. The arrival of author identification initiatives like ORCID [13] and online research profiles such as ResearchGate or Google Scholar should make it easier to find working contact information for authors in the future.

Considering only the papers from 2011, our results show that asking authors for their data shortly after publication does yield a moderate proportion of datasets (c. 40%). A comparable study [11] received 59% of the requested datasets from papers that were less than a year old. It is hard to tell whether this difference is due to the slightly different research communities involved or the presence of an extra year between publication and the data request in this study. A related paper by Wicherts et al. in 2005 [9] received only 26% of requested psychology datasets.

Overall, we only received 19.5% of the requested datasets, and only 11% for articles published before 2000. We found that several factors contribute to these low proportions: non-working emails, a 50% response rate, and sometimes the lack of an informative response from the authors. However, when the authors did give the status of their data, the proportion of datasets that still existed dropped from 100% in 2011 to 33% in 1991 (Figure 1D). Unfortunately, many of these missing datasets could be retrieved only with considerable effort by the authors, and others are completely lost to science.

Many datasets produced in scientific research are unique to their time and location, and once lost they cannot be replaced [14]. Since it is impossible to know what uses would have been found for these data, or when they would become important, leaving their preservation to authors denies future researchers any chance of reusing them. Fortunately, one effective solution is to require that authors share it on a public archive at publication: the data will be preserved in perpetuity, and can no longer be withheld or lost by authors. Some journals have already enacted policies to this effect [e.g. 5,6], and we hope that the worrying magnitude of the issues reported here will encourage others to draft similar policies in due course.

## Experimental Procedures

It is likely that expectations on data sharing will differ between academic communities, and that some data types are easier to preserve than others. Moreover, the types of data being collected change through time. We attempted to control for these effects by focusing on a single type of data that has been collected in the same way for many decades: data on morphological dimensions from plants or animals, as is typically collected by biologists and taxonomists. We are also conducting a parallel study on how the reproducibility of statistical analyses changes through time, and this study is working on reproducing discriminant function analyses (DFA), which are commonly applied to morphometric data [15]. We therefore also set the condition that the data must have been used in a DFA.

We searched Web of Science for articles matching 'morpholog* and discriminant' in the topic field for the years 1980 to 2011. Only 24 papers were identified before 1991, and these were excluded. To reduce the total number of articles, we chose to focus on odd years from 1991 to 2011, leaving 1009 papers. These papers were randomly assigned to the working group for data collection. Papers were excluded if the article text was not available to us either online or via the University of British Columbia library, if the analysis did not include morphological data from a biological organism, or if the paper did not report the results of a DFA. Papers were also excluded if the data were already available as a supplementary file, appendix, or on another website, as curation of these datasets is no longer the responsibility of the author. Due to the effort involved in checking all 1009 papers for details on analysis and author contact information, we stopped data collection after a random subset of 526 papers had been assessed. Of these, 10 did not meet the inclusion criteria (e.g. were not DFAs on morphology, or had data already available in a supplementary file or appendix), and were dropped. This left 516 papers, with a minimum of 26 papers for any given year, and over 40 for most years (Table 1). Interestingly, we found that only 2.4% (13 of 529) of otherwise eligible papers had made their data available at publication: one paper each from 1999, 2001, 2003 and 2007, three papers in 2005, two in 2009, and four in 2011.

Data collected from the papers included information on the DFA used and results (for the reproducibility analyses), and author contact information. In every case, we attempted to find email addresses for the first, corresponding, and last authors of every paper. Often these were not mutually exclusive (e.g. a single author), and there were many different combinations. We attempted to extract the emails from the article text, but quickly determined that older papers would be more likely to have non-working email addresses [12] or no emails at all. We therefore also searched online for a maximum of five minutes per author for a recent or current email address.

We used R [16] to generate data request emails, with all available email addresses in the 'to:' field, and used an R script to automatically send them out on April 15th, 2013. Reminder emails were sent out to unresponsive authors three weeks later (May 8th, 2013). When authors replied asking for more information, we provided additional details as required. The text for these two emails is included in the Supplemental Material. The recording period for author responses ended on the 5th of June, 2013, and the papers sorted into different outcomes: 1) all email addresses generated an error message, 2) no response received, 3) a response was received but gave no information about the status of the data, 4) data lost or stored on obsolete hardware, and 5) the authors had the data but were unwilling to share, or 6) data received. Since outcomes (5) and (6) both implied that the dataset still existed, we combined these into a single outcome 'data extant'.

We used logistic regression to investigate the relationship between the age of a paper and the probability that the data were still extant. Further sub-analyses were conducted on subsets of the data to investigate the relationships between the age of the paper and 1) the probability that at least one email appeared to work; 2) the conditional probability of a response given that at least one email appeared to work; 3) the conditional probability of getting a response that indicated the status of the data (data lost, exist but unwilling to share, or data shared) given that a response was received; and finally 4) the conditional probability the data was extant given that an informative response was received. We also used Poisson regressions to investigate the relationship between article age and the number of emails found in the paper or online. Lastly, logistic regressions were used to examine how article age affected the chance that an email address appeared to work. All analyses were carried out in R 3.0.1 [16]; the analysis code and data are available on Dryad (doi:10.5061/dryad.q3g37).

## Acknowledgements

We thank M. Whitlock, M. O'Connor, E. Hart, E. Wolkovich and T. Veen for helpful discussions, and two anonymous reviewers for comments on an earlier version. KJG and DJR were supported by NSERC postgraduate studentships, DGB by an NSERC Vanier CGS and a Killam doctoral scholarship, FD by an NSERC CREATE Training Program in Biodiversity Research), and J-SM by a Fond Québécois de la Recherche sur la Nature et les Technologies (FQRNT) postdoctoral scholarship. SR was supported by an NSERC postdoctoral fellowship and MTF by the Investment Agriculture Foundation. Lastly, we are very grateful to the researchers we contacted for their contributions to the data presented here.

Table 1. Breakdown of data availability by year of publication. Data are displayed as N [%]; the percentages are calculated by rows.

| Year | no working email | no response to email | response did not give status of data | data lost | data exist, unwilling to share | data received | data extant (unwilling to share + received) | Number of papers |
|---|---|---|---|---|---|---|---|---|
| 1991 | 9 [35%] | 9 [35%] | 2 [8%] | 4 [15%] | 1 [4%] | 1 [4%] | 2 [8%] | 26 |
| 1993 | 14 [39%] | 11 [31%] | 3 [8%] | 7 [19%] | 0 [0%] | 1 [3%] | 1 [3%] | 36 |
| 1995 | 11 [31%] | 9 [26%] | 0 [0%] | 7 [20%] | 2 [6%] | 6 [17%] | 8 [23%] | 35 |
| 1997 | 11 [37%] | 9 [30%] | 1 [3%] | 2 [7%] | 3 [10%] | 4 [13%] | 7 [23%] | 30 |
| 1999 | 19 [48%] | 13 [32%] | 1 [2%] | 1 [2%] | 0 [0%] | 6 [15%] | 6 [15%] | 40 |
| 2001 | 13 [30%] | 15 [35%] | 3 [7%] | 4 [9%] | 0 [0%] | 8 [19%] | 8 [19%] | 43 |
| 2003 | 9 [20%] | 20 [43%] | 4 [9%] | 2 [4%] | 0 [0%] | 11 [24%] | 11 [24%] | 46 |
| 2005 | 11 [24%] | 14 [31%] | 6 [13%] | 1 [2%] | 0 [0%] | 13 [29%] | 13 [29%] | 45 |
| 2007 | 12 [18%] | 31 [47%] | 2 [3%] | 4 [6%] | 1 [2%] | 16 [24%] | 17 [26%] | 66 |
| 2009 | 9 [13%] | 34 [49%] | 3 [4%] | 5 [7%] | 6 [9%] | 12 [17%] | 18 [26%] | 69 |
| 2011 | 13 [16%] | 29 [36%] | 8 [10%] | 0 [0%] | 7 [9%] | 23 [29%] | 30 [38%] | 80 |
| Totals: | 131 [25%] | 194 [38%] | 33 [6%] | 37 [7%] | 20 [4%] | 101 [19%] | 121 [23%] | 516 |

Figure 1. The effect of article age on four obstacles to receiving data from the authors. A) Predicted probability that the paper had at least one apparently working email. B) Predicted probability of receiving a response, given that at least one email was apparently working. C) Predicted probability of receiving a response giving the status of the data, given that we received a response. D) Predicted probability that the data were extant (either 'shared', or 'exist but unwilling to share') given that we received a useful response. In all panels, the line indicates the predicted probability from the logistic regression, the grey area shows the 95% CI of this estimate and the red dots indicate the actual proportions from the data.

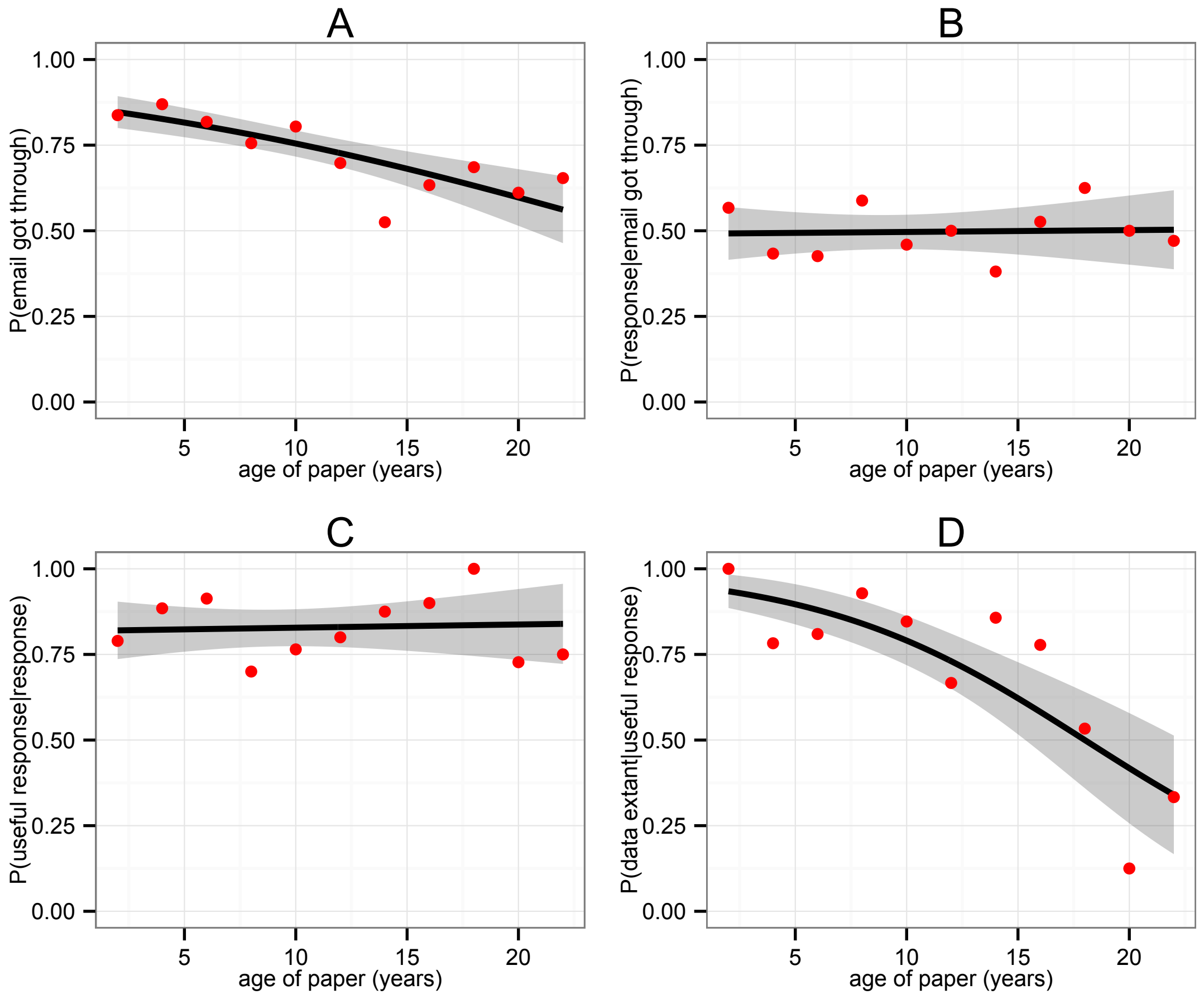

Figure 2. The effect of article age on the number and status of author emails. A) Number of emails found in the paper against article age. B) Predicted probability that an individual email from the paper appeared to work against article age. C) Number of emails found by searching on the web against article age. D) Predicted probability that an individual email found on the web appeared to work against article age. The line indicates the predicted probability from a Poisson (A, C) or logistic (B, D) regression, the grey area shows the 95% CI of this estimate and the red dots indicate the actual proportions from the data.

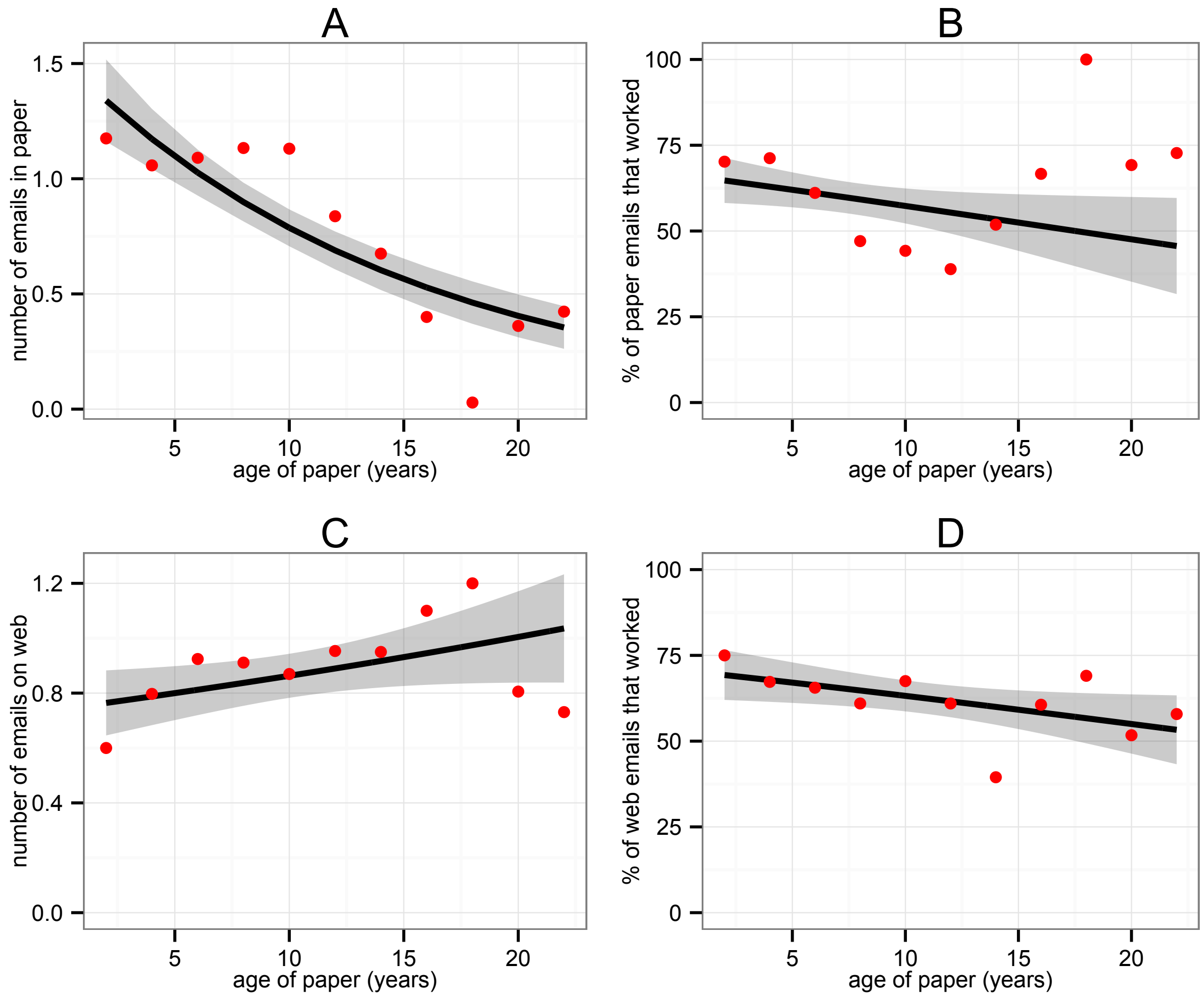